\documentclass[letterpaper]{article} 
\usepackage[]{aaai2027}  
\usepackage[hyphens]{url}  
\usepackage{graphicx} 
\usepackage{natbib}  
\usepackage{caption} 
\usepackage{algorithm}
\usepackage{algorithmic}

\usepackage{newfloat}
\usepackage{listings}
\DeclareCaptionStyle{ruled}{labelfont=normalfont,labelsep=colon,strut=off} 
\floatstyle{ruled}
\newfloat{listing}{tb}{lst}{}
\floatname{listing}{Listing}

\usepackage{booktabs}

\usepackage{dsfont}
\usepackage{multirow}
\usepackage{tabularx}
\usepackage{array}
\usepackage{bm}
\usepackage{threeparttable}
\usepackage[table,dvipsnames,svgnames]{xcolor}
\usepackage{colortbl}
\usepackage{subcaption}

\usepackage[T1]{fontenc}
\usepackage[utf8]{inputenc}

\usepackage{xurl}
\usepackage{amsmath}
\usepackage{amssymb}
\usepackage{mathtools}
\usepackage{amsthm}

\title{Protecting Creative Writing Copyright against AI Imitation via Implicit Watermarking}
\author{
    Ziwei Zhang\textsuperscript{\rm 1},
    Juan Wen\textsuperscript{\rm 1}\corresponding,
    Wanli Peng\textsuperscript{\rm 1}\corresponding,
    Zhengxian Wu\textsuperscript{\rm 1},
    Yinghan Zhou\textsuperscript{\rm 1},
    Yiming Xue\textsuperscript{\rm 1}
}
\affiliations{
    \textsuperscript{\rm 1}China Agricultural University\\

}

\begin{document}

\maketitle

\begin{abstract}
Large language models (LLMs) enable powerful knowledge injection through approaches such as in-context learning and fine-tuning, but they also introduce new risks of unauthorized imitation of high-value creative works. Existing copyright protection techniques mainly focus on visual media, leaving the protection of creative writing largely unexplored. In this work, we investigate a new challenge: verifying whether AI-generated texts inherit the creative essence of protected works without authorization. We propose WIND (Watermarking via Implicit and Non-disruptive Disentanglement), a zero-watermarking framework that constructs an implicit and verifiable creative signature for copyright verification. WIND decomposes creative essence into five complementary dimensions and leverages an LLM-based instance delimitation mechanism to extract condensed representations of protected creative characteristics. By disentangling creative-specific information from irrelevant textual variations, these representations are mapped into a compact watermark space without modifying the original texts. Extensive experiments demonstrate that WIND achieves over 98\% F1 scores while maintaining low false-positive rates, substantially outperforming existing watermarking and text classification approaches under various AI imitation scenarios.
\end{abstract}


\section{Introduction}

Large language models (LLMs) utilize some efficient knowledge injection approaches to achieve significant performance gains across downstream tasks. These techniques primarily follow three paradigms: In-context learning for update-free knowledge acquisition \cite{dong-etal-2024-survey,brown2020language, wei2022chain, liu2023pre, liu2024adaptive, DBLP:journals/corr/abs-2303-08774}, parameter-efficient fine-tuning for minimal-adjustment adaptation \cite{DBLP:conf/iclr/HuSWALWWC22, liu2021p, han2024parameter}, and knowledge editing for precise factual modification \cite{feng2023trends,wang2024knowledge}.  These techniques have heightened concerns over copyright infringement of authors’ creative works, resulting in a growing number of legal disputes. In the domain of creative writing, for instance, \textit{Authors Guild of America} sued \textit{OpenAI} for training its models on copyrighted texts without permission \cite{AIauthors}. Similarly, \textit{The New York Times} also filed a lawsuit against \textit{OpenAI} \cite{Times}. Similar litigation has arisen in other creative domains, such as visual arts \cite{Sarah,getty2023stability}. Consequently, the protection of creative works has drawn significant attention from researchers \cite{liu2023watermarking, tang2023did, maini2024llm}.

\begin{figure}[t]
    \centering
    \includegraphics[scale=0.3]{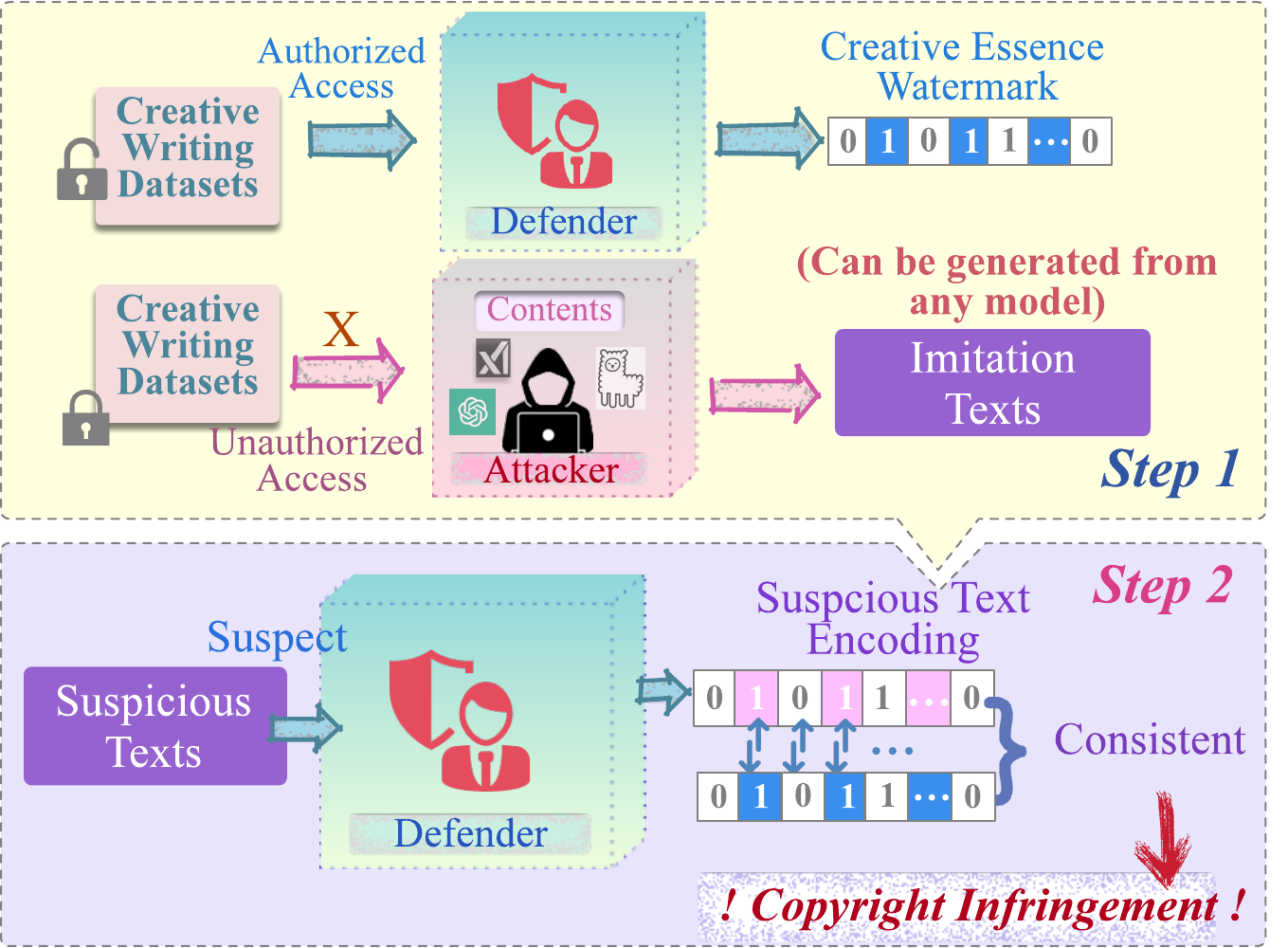} 
    \caption{The application scenario of the implicitly verifiable watermark.}
    \label{process}
\end{figure}

Current methods for ensuring copyright protection in creative works primarily focus on digital watermarking, a widely studied and validated paradigm for safeguarding data and preventing infringement. Several studies \cite{9859698, salman2023raising, shan2023glaze} have explored scrambled watermarks, which involve embedding intentional signals into images to protect authors' creative visual arts. Alternatively, recent work \cite{huangdisentangled} explores verifiable watermarks using diffusion models to establish clear copyright boundaries for image style protection. 
Current watermarking techniques are predominantly tailored for visual media. However, effective methodologies for creative writing remain significantly underexplored.

To defend against unauthorized imitation by LLMs, we aim to construct a verifiable watermark that enables copyright verification for creative writing. Specifically, our goal is to determine whether an AI-generated text has inherited the protected creative characteristics of a given work without authorization. Different from authorship identification, which focuses on inferring the likely source writer of a given text, our objective is to proactively protect creative works by establishing an implicit and verifiable creative signature. This shifts the problem from passive text attribution to active copyright verification, where the core challenge lies in capturing and validating the underlying creative essence rather than identifying the author of a single text.

Crucially, copyright verification requires more than passive similarity estimation between texts. Instead, it demands an implicit mechanism that captures the underlying creative essence of protected works and transforms it into a verifiable representation. Drawing on linguistic studies of writing characteristics \cite{vaezi2019development,tweedie1998variable,lu2010automatic,steen2010method,pennebaker1999linguistic,kao2012computational}, we define creative writing signature through five key elements: (1) vocabulary and word choice (VWC), (2) syntactic structure and grammatical features (SSGF), (3) rhetorical devices and stylistic choices (RDCS), (4) tone and sentiment (TS), and (5) rhythm and flow (RF). These elements collectively characterize the creative essence of writing beyond surface-level textual similarity. Based on this representation, we propose WIND, which disentangles creative-specific information from irrelevant variations and encodes the extracted creative signature into an implicit and verifiable watermark for copyright verification.


Regarding prior knowledge selection's impact on LLM parsing \cite{leidinger2023language}, we introduce dual prompt templates, contrastive learning, and an instance delimitation mechanism to select optimal contexual references for each sample. To disentangle creative-specific from irrelevant features, we employ LLMs to extract the five elements to establish the creative writing essence, termed \textit{condensed-lists}. These condensed representations are further projected into a disentangled space, where a stable creative signature anchor is derived and implicitly encoded as a verifiable watermark.
Figure \ref{process} illustrates the WIND application scenario. WIND (the Defender) generates a unique watermark from creative-specific features extracted from protected creation. For a suspicious AI-generated text, the defender validates whether it inherits the protected creative essence by measuring the distance between its watermark representation and the protected watermark anchor. WIND is optimized for few-shot scenarios to reduce computational costs for practical applications.  Code is available in the supplements. Our key contributions include:
\begin{itemize}
\item We present WIND, a novel implicit watermarking framework that enables copyright verification of creative writing against unauthorized AI imitation. To the best of our knowledge, this is the first work to formally decompose the abstract essence of creative writing and leverage it for copyright protection via watermarking.
\item We create an instance delimitation mechanism to identify optimal prior knowledge, which facilitates the extraction of condensed-lists by LLMs. Subsequently, we establish a verifiable watermark domain for creative essence, moving beyond explicit signal injection methods that may alter fragile textual characteristics.
\item Extensive experiments confirm the method's effectiveness and robustness against a wide range of challenges, including state-of-the-art text watermarking techniques, robust attacks, and multi-level visual analysis.
\end{itemize}

\begin{figure*}
    \centering
   
    \includegraphics[scale=0.43]{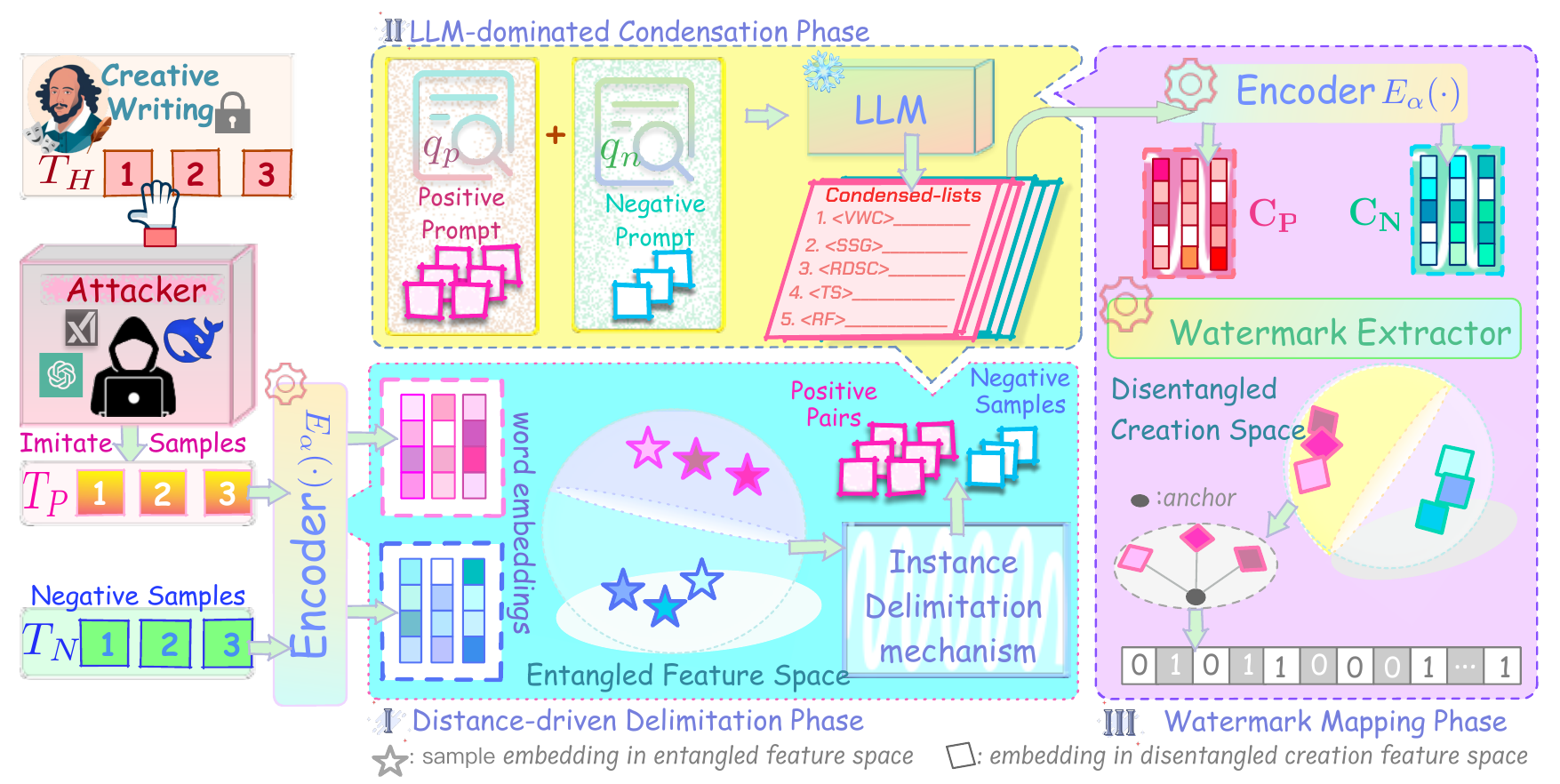}
    \caption{The overall framework of WIND, which consists of three main phases: (1) Distance-driven Delimitation, (2) LLM-dominated Condensation, and (3) Watermark Mapping. The numbers within the squares of $\bm{T_H}$, $\bm{T_P}$, and $\bm{T_N}$ represent different samples. Additionally, a star in the entangled feature space and a diamond in the disentangled space (of the same color) denote the same sample.}
     \label{structure}
\end{figure*}

\section{Related Work}
\subsection{Creative Work Copyright Protection}
Existing methods for detecting infringements of creative works mainly fall into two categories: membership inference (MI) and digital watermarking. For MI-based approaches, \cite{shi2024detecting} compares data generated before and after model training, while \cite{maini2024llm} performs membership inference attacks (MIAs) in gray-box settings. However, these methods struggle when LLMs imitate the unique creation but modify irrelevant details, and gray-box model access limits real-world applications.

Digital watermarks for copyright protection are primarily categorized into scrambled and verifiable types. Scrambled methods protect visual arts by embedding signals through reversible transformations \cite{9859698}, adversarial perturbations \cite{salman2023raising}, or cloaks \cite{shan2023glaze}, although they remain vulnerable at the latent representation level. Conversely, verifiable watermarks offer robust defense against unauthorized use, as demonstrated in image style protection for generative models \cite{huangdisentangled}. Despite these visual advances, safeguarding creative writing remains a critical and underexplored challenge.

\subsection{Text watermarking for LLMs}
Text watermarking protects LLM creative styles \cite{he2022protecting} by modifying token sampling or internal logic during generation. Existing approaches utilize random sequences \cite{christ2024undetectable}, Levenshtein distance \cite{kuditipudirobust}, green-list biasing \cite{kirchenbauer2023watermark, zhaoprovable}, high-entropy weighting \cite{lu-etal-2024-entropy}, introducing unbiased attribution \cite{huunbiased}, and adaptively adjusting watermark strength \cite{wang-etal-2025-morphmark}. However, these methods fail to safeguard the creative ideas. We propose WIND, extracting creative essence to generate implicit watermarks, ensuring verifiable copyright protection without distorting fragile creative attributes.

\section{Approach}
\label{sec:3}
\subsection{Problem Formulation}
\label{3.1}
\textbf{Creative Writing Essence.} We define the protected creative writing ($\bm{T_H}$) as a collection of human-authored works sharing a common creative signature. Next, we operationalize the abstract concept of "creative writing essence" by decomposing it into five concrete elements \cite{vaezi2019development,tweedie1998variable,lu2010automatic,steen2010method,pennebaker1999linguistic,kao2012computational}, including vocabulary and word choice (VWC), syntactic structure and grammatical features (SSGF), rhetorical devices and stylistic choices (RDCS), tone and sentiment (TS), and rhythm and flow (RF). See supplements for details. 

\textbf{Attackers.} Attackers are assumed to have access to publicly available or unauthorizedly obtained creative works, which can be used as references for generating imitation texts. Furthermore, attackers may employ LLM APIs or other black-box generation systems, making the details of their imitation process unavailable to the defender.

\textbf{Defender.} Our defense objective is to verify whether a suspicious AI-generated text inherits the protected creative essence without authorization. The defender $D(\cdot)$ generates an implicit watermark representation for protected creative writing. To mitigate statistical biases between human-written and machine-generated texts, we construct two sets of generated texts: TP, which imitates the protected creative writing $\bm{T_H}$, and $\bm{T_N}$, which represents unrelated writing. For a suspicious text $\bm{T_{test}}$, we extract its watermark $D(\bm{T_{test}})$ and measure its Hamming distance to the predefined protected watermark anchor. The verification score is defined as:
\begin{equation}
pr = 
\begin{cases} 
1, & \text{if } d_h(D(\bm{T}_{\text{test}}), D(\bm{T}_P)) < \epsilon \\
0, & \text{otherwise}
\end{cases}
\end{equation}
where $d_h$ denotes Hamming distance and $\epsilon$ empirically is 1\% of the length of watermark.

\subsection{Overview}
\label{overview}

The training process of WIND is depicted in Figure \ref{structure}. Our pipeline operates through three sequential phases designed to progressively isolate and encode the creative essence. The process begins with a \textit{distance-driven delimitation phase}, which constructs a feature space to identify samples that closely emulate the protected creative style from those that do not, thus preparing an optimal contextual prior for subsequent distillation. This is followed by an \textit{LLM-dominated condensation phase}, which is the core of disentanglement approach. Here, we leverage the powerful generative prior of an LLM, conditioned on the previously identified context, to interpret and distill the abstract creative essence into a structured set of concrete stylistic elements. Finally, the \textit{watermark mapping phase} transforms this structured stylistic information into a compact and verifiable zero-watermark, ensuring robust copyright verification.

\subsection{Distance-driven Delimitation Phase}
\label{phase1}

We employ the encoder with $de$ dimensions and adjustable parameters $\alpha$, denoted as $E_\alpha(\cdot)$, to compute word embeddings for $\bm{T_P}$ and $\bm{T_N}$. Each sentence $t_p{_{i}}\in \bm{T_P}$ and $t_n{_{j}}\in \bm{T_N}$ is mapped into a positive feature vector $\bm{p_i} \in \mathbb{R}^{b_i \times de}$ and a negative feature vector $\bm{n_j} \in \mathbb{R}^{b_j \times de}$, respectively, with $b_i$ and $b_j$ representing the number of words in $t_p{_{i}}$ and $t_n{_{j}}$. 
Assuming both $\bm{T_P}$ and $\bm{T_N}$ contain $num$ samples, their corresponding feature vector sets are denoted as $\bm{P} = [\bm{p_1}, \bm{p_2}, \dots, \bm{p_{num}}]$ and $\bm{N} = [\bm{n_1}, \bm{n_2}, ..., \bm{n_{num}}]$, respectively. Both $\bm{P}$ and $\bm{N}$ inherently include creation-irrelevant features. In this context, texts from the $\bm{T_P}$ are considered positive, while all other texts are negative. 
Next, for any feature vector $\bm{x} \in \bm{P} \cup \bm{N}$, we use the cosine similarity function $sim(\cdot)$ to identify the most similar vector to $\bm{x}$ from 
the union of $\bm{P}$ and $\bm{N}$, i.e., $\bm{y_x^*} = \operatorname{arg\,max}_{\bm{y} \in \bm{P} \cup \bm{N} \backslash \{\bm{x}\}} sim(\bm{x}, \bm{y})$, with the highest similarity expressed as $sim_x^* = sim(\bm{x}, \bm{y_x^*})$. 
Then the cross-entropy loss $\mathcal{L}_{ce}$ is then computed as:
\begin{equation}
\mathcal{L}_{ce}
=
-\frac{1}{2 \times num}
\sum_{\bm{x} \in \bm{P} \cup \bm{N}}
\log p_{x,y_x}.
\label{ce}
\end{equation}
where $y_x$ is ground-truth of sample $\bm{x}$. $\hat{y}_x$  is the pseudo-label determined by the class of the most similar vector $\bm{y_x^*}$. Specifically, $\hat{y}_x=1$ holds when $\bm{y_x^*} \in \bm{P}$, otherwise, $\hat{y}_x=0$.
Moreover, to emphasize the distinctions between positive and negative samples, we design a hyperparameter $mar$ and
utilize a contrastive loss function:
\begin{equation}
\label{con}
\begin{split}
\mathcal{L}_{con} = \frac{1}{2\times num} ( \sum_{\bm{x}, \bm{x'} \in \bm{P}} \|\bm{x} - \bm{x'}\|^2 +\\ \sum_{\bm{x} \in \bm{N}, \bm{x''} \in \bm{P}} \text{max}(0,mar-\|\bm{x} - \bm{x''}\|^2 )).
\end{split}
\end{equation}
The importance of optimal reference instance selection in prompt engineering has been empirically established \cite{sahoo2024systematic}. Taking this viewpoint, we introduce an instance delimitation mechanism to select the optimal prior knowledge for each sample. For each $\bm{x}$, the most similar vector $\bm{y_x^*}$ may come from either set $\bm{P}$ or $\bm{N}$. We construct two sets: one is the positive pair set $\bm{pp}$, and the other is the negative sample set $\bm{neg}$. Specifically, $\bm{pp}$ contains samples that closely emulate $\bm{T_H}$, where the most similar sample (with similarity exceeding a predefined threshold) is labeled as the positive instance. Each sample in $\bm{pp}$ is paired with its corresponding optimal prior knowledge $\bm{y_x^*}$, allowing better disentanglement of the creation-specific features that distinguish protected creative works from unprotected ones. In contrast, $\bm{neg}$ is composed of individual samples instead of pairs due to the diverse creative works in $\bm{T_N}$, whereas $\bm{T_P}$ sentences uniformly exhibit the creative essence.
The assignments for $\bm{pp}$ and $\bm{neg}$ are formalized in the corresponding equations:
\begin{equation}
 \label{pairs}
	\bm{pp} = \{(\bm{x}, \bm{y_x^*}) \mid \bm{x} \in \bm{P} \cup \bm{N} \land \bm{y_x^*} \in \bm{P} \land sim_x^* > \sigma\},
 \end{equation}
 \begin{equation}
 \label{neg}
 \bm{neg} = \{\bm{x} \mid \bm{x} \in \bm{P} \cup \bm{N} \land (\bm{y_x^*} \in \bm{N} \vee sim_x^* \leq \sigma)\} ,
 \end{equation}
where $\sigma$ is a predefined threshold and $sim_x^*$ denotes the similarity of $\bm{x}$ to its nearest neighbor. We adopt an instance delimitation mechanism instead of direct label prediction via $E_{\alpha}(\cdot)$ to mitigate misclassification.

\subsection{LLM-dominated Condensation Phase}
\label{phase2}

To further disentangle the creative essence, we employ an LLM to extract condensed-lists composed of five elements. Specifically, we design two prompt templates, $q_p$ and $q_n$, where $q_p$ is designed for samples in the positive pair set $\bm{pp}$, and $q_n$ is used for the negative sample set $\bm{neg}$. For each sample $\bm{t_m} \in \bm{T_P} \cup \bm{T_N}$, we construct the input sequence by combining the prompt with either the pair $(\bm{t_m}, \bm{t_s})$ or the single sample $\bm{t_m}$. Here, $q=q_p$ and the input includes its most similar neighbor $\bm{t_s}$ if they form a pair in $\bm{pp}$, otherwise $q=q_n$ for the samples in $\bm{neg}$. The concatenated input $q_p||(\bm{t_m},\bm{t_s})$ or $q_n||\bm{t_m}$ is then fed into a frozen-parameter LLM, denoted as $G(\cdot)$, which generates a condensed-list $\bm{c} = [e_1, e_2, \ldots, e_5]$ representing five key elements per sample. Prompt details are in supplements.

For each condensed-list $\bm{c}_m$, we introduce $\bm{z}_m = E_{\alpha}(\bm{c}_m)$, then we get $\bm{C}_P = \{\bm{z}_m \mid \bm{t_m} \in \bm{T}_P\}$ and $\bm{C_N} = \{\bm{z_m} \mid \bm{t_m} \in \bm{T}_N\}$. 
To separate positive and negative samples and obtain more discriminative watermarks, we impose a contrastive objective on condensed-list embeddings. Its first term clusters embeddings from $\bm{T}_P$, while the second penalizes embeddings from $\bm{T}_N$ whose cosine similarity to positive ones exceeds margin $\rho$, where $\tilde{\bm{z}}_i$ and $\tilde{\bm{z}}_j$ are L2-normalized representations.
\begin{equation}
\label{lstyle}
\begin{split}
 \mathcal{L}_{style}=\frac{1}{|\bm{C}_P|(|\bm{C}_P|-1)}\sum_{\substack{\bm{z}_i,\bm{z}_j \in \bm{C}_P \\ i \neq j}}\left(1 -\tilde{\bm{z}}_i^{\top}\tilde{\bm{z}}_j\right)+ \\
    \frac{1}{|\bm{C}_N||\bm{C}_P|}\sum_{\bm{z}_i\in \bm{C}_N}\sum_{\bm{z}_j \in \bm{C}_P}\max\left(0,\tilde{\bm{z}}_i^{\top}\tilde{\bm{z}}_j - \rho\right).
\end{split}
\end{equation}

\subsection{Watermark Mapping Phase}
\label{phase3}
These lists are then further transformed into positive disentangle embeddings $\bm{C_P}$ and negative embeddings $\bm{C_N}$ through the encoder $E_ \alpha(\cdot)$. 
It is worth noting that this encoder is identical to the one used in the first step. We then employ the sigmoid function $\theta (\cdot)$ and a learnable watermark matrix $\bm{M}_\gamma \in \mathbb{R}^{len \times de}$ to construct the watermark extractor, where $\gamma$ denotes the learnable parameters and $len$ is the fixed watermark length. Each condensed-list $\bm{c_m}$ is processed according to the formula:
\begin{equation}
\label{w}
    \bm{w_m}= \theta(\bm{M}_\gamma \cdot E_ \alpha(\bm{c_m})),
\end{equation}
where $\bm{w_m}\in \bm{W}$ and $\bm{W} \in \mathbb{R}^{2num \times len}$. We define a fixed secret anchor $\bm{a} =[\bm{a}_1, \bm{a}_2,...,\bm{a}_{len}]$, where $\bm{a}\in\{0,1\}^{len}$. Its complement $\bar{\bm{a}} = \mathbf{1} - \bm{a}$ is used as the negative watermark target for unprotected texts. This avoids the instability caused by batch-dependent anchor estimation. 
We anticipate that all samples imitating $\bm{T_H}$, after being mapped by $\bm{M}_\gamma$, will closely converge in a disentangled creative feature space. 

\subsection{Training Procedure}
\label{training}
After generating condensed-lists for all instances in $\bm{T_{P}}\cup \bm{T_{N}}$, their watermark vectors are partitioned according to their original source sets. Specifically, $\bm{W_{P}}$ contains watermarks extracted from imitation texts $\bm{T_{P}}$, while $\bm{W_N}$ contains watermarks extracted from unprotected texts $\bm{T_N}$. The prompt selection is still determined by $\bm{pp}$ and $\bm{neg}$,  whereas watermark supervision is assigned according to whether the original sample belongs to $\bm{T_{P}}$ or $\bm{T_N}$.

To generalize the creative essence, the positive watermark loss $L_w^{+}=\frac{1}{|\bm{W_P}|}\sum_{\bm{w_i} \in \bm{W_P}}BCE(\bm{w_i}, \bm{a})$ encourages watermarks extracted from $\bm{T_{P}}$ to match the fixed anchor $\bm{a}$, while the negative watermark loss $L_w^{-}=\frac{1}{|\bm{W}_N|}\sum_{\bm{w}_j \in \bm{W}_N}BCE(\bm{w}_j, \bar{\bm{a}})$ pushes $\bm{T_N}$ toward the complement anchor $\bar{\bm{a}}$, where $BCE$ is Binary Cross-Entropy loss. This explicit positive-negative supervision prevents negative samples from collapsing to the protected watermark region. The final watermark loss is calculated as: 
\begin{equation}
\label{wmbce}
\mathcal{L}_w=\lambda_{w}^{+} L_w^{+}+\lambda_{w}^{-} L_w^{-},
\end{equation}
where $\lambda_{w}^{+}$ and $\lambda_{w}^{-}$ are 
hyperparameters. $\mathcal{L}_{w}$ encourages all samples in $\bm{T_{P}}$ to converge towards the representative anchor $\bm{a}$ derived from $\bm{pp}$. Accordingly, the total loss for WIND is:
\begin{equation}
\label{loss}
\mathcal{L}=\mathcal{L}_{ce}+\mathcal{L}_{con}+\lambda_{c} \mathcal{L}_{style}+\mathcal{L}_w.
\end{equation}
The training procedure is shown in Algorithm 1.

\begin{algorithm}[t]
  \caption{Training Procedure of WIND}
  \label{alg:wind_training}
  \begin{algorithmic}
    \STATE {\bfseries Input:} Protected creative writing $\bm{T_H}$, imitation texts $\bm{T_P}$, unprotected texts $\bm{T_N}$, encoder $E_\alpha(\cdot)$, similarity function $sim(\cdot)$, watermark matrix $\bm{M}_\gamma$, sigmoid $\theta(\cdot)$, LLM $G(\cdot)$, prompts $q_p$, $q_n$, $R$ episodes and $ep$ epochs.
    \STATE {\bfseries Output:} Updated parameters $\alpha$, $\gamma$ and anchor $\bm{a}$.
    
    \STATE Initialize a fixed anchor $\bm{a} \in \{0,1\}^{len}$ and its complement $\bar{\bm{a}}=\bm{1}-\bm{a}$.
    
    \FOR{$epoch = 1$ {\bfseries to} $ep$}
      \FORALL{$episode \in R$}
        \FORALL{$\bm{t_m} \in \bm{T_P} \cup \bm{T_N}$}
          \STATE $\bm{x_m} = E_\alpha(\bm{t_m})$
          \STATE $\bm{y^*_{x_m}} = \mathrm{argmax}_{\bm{y} \in \bm{P} \cup \bm{N} \setminus \{\bm{x_m}\}} sim(\bm{x_m}, \bm{y})$
        \ENDFOR
        
        \STATE Construct $\bm{pp}$ and $\bm{neg}$ using Eq. \ref{pairs} and Eq. \ref{neg}.
        
        \FORALL{$\bm{t_m} \in \bm{T_P} \cup \bm{T_N}$}
          \STATE Select prompt $q_p$ or $q_n$ according to whether $(\bm{x_m},\bm{y^*_{x_m}}) \in \bm{pp}$ or $\bm{x_m} \in \bm{neg}$.
          \STATE Generate condensed-list $\bm{c_m}$ with $G(\cdot)$.
          \STATE Compute watermark $\bm{w_m} = \theta(\bm{M}_\gamma \cdot E_\alpha(\bm{c_m}))$.
        \ENDFOR
        
        \STATE Split watermarks into $\bm{W_P}$ and $\bm{W_N}$ according to $\bm{T_P}$ and $\bm{T_N}$.
        \STATE Compute $\mathcal{L}_{ce}$, $\mathcal{L}_{con}$, $\mathcal{L}_{style}$, $\mathcal{L}_{w}$.
        \STATE Update $\alpha,\gamma$ with overall loss $\mathcal{L}$ in Eq. \ref{loss}.
      \ENDFOR
    \ENDFOR
  \end{algorithmic}
\end{algorithm}

\subsection{Watermark Validation}
\label{val}
The goal of watermark validation is to determine whether a suspicious text inherits the protected creative essence associated with the watermark. During testing, upon receiving the input sentence $\bm{t_{test}}$, we identify the most similar sample $\bm{t_{ts}}$ from the training dataset. We then extract the condensed list $\bm{c_{test}}$ by leveraging the frozen LLM $G(\cdot)$ with the input $(\bm{t_{test}}, \bm{t_{ts}})$ or $\bm{t_{test}}$. The $\bm{c_{test}}$ is formulated as follows:
\begin{equation}
\label{val1}
\bm{c_{test}}= G(q||(\bm{t_{test}},\bm{t_{ts}})) \text{ or } G(q||\bm{t_{ts}}).
\end{equation}
Here, $q$ is the prompt template that depends on classification of $\bm{t_{test}}$ as either $\bm{pp}$ (paired with $\bm{t_{ts}}$) or $\bm{neg}$, following the mechanism mentioned before.
Subsequently, $\bm{c_{test}}$ is mapped into the disentangled feature space, facilitating the extraction of unique creation features represented as $\bm{w_{test}} = \theta(\bm{M}_\gamma \cdot E_\alpha(\bm{c_{test}}))$. This process quantifies the similarity that the tested sample $t_{test}$ imitates the protected creative writing, formulated as follows: 
\begin{equation}
\label{acc}
\mathcal{P}(\bm{w}_{test}|\bm{a})=\frac{1}{len}\sum_{i=1}^{len}\mathbb{I}(r(\bm{w}_{test}^{i}) = a_i).
\end{equation}
Let $\mathbb{I}(\cdot)$ denote the indicator function, where  $\mathbb{I}(\cdot) = 1$ if $r(\bm{w}_{test}^i) = a^i$. $r(\cdot)$ acts as a hard thresholding function, converting its input into a bit string. where For robust copyright verification, $\mathcal{P}$ approaches 1 when $t_{test}$ imitates creative writing and 0 otherwise. The validation algorithm is shown in supplements.

\begin{table*}[t]
\renewcommand{\arraystretch}{0.7}
\centering
\begin{tabular}{c l c c c c c c}
\toprule
\multirow{2}{*}{\textit{num}} & \multirow{2}{*}{Methods}
& \multicolumn{3}{c}{ROC}
& \multicolumn{3}{c}{SP} \\
\cmidrule(lr){3-5} \cmidrule(lr){6-8}
& & F1 & TPR & FPR & F1 & TPR & FPR \\
\midrule
\multirow{6}{*}{10}
& BERT     & 71.67(3.81) & 64.05(4.27) & 14.69(7.80) & 74.25(6.81) & 90.75(2.53) & 53.68(9.82) \\
& RoBERTa  & 82.92(2.58) & 87.32(3.78) & 23.29(4.34) & 78.68(4.54) & 89.59(5.72) & 38.13(5.28) \\
& T5       & 63.21(7.69) & 68.75(4.52) & 48.79(3.90) & 66.83(4.84) & 58.08(8.59) & 15.73(1.67) \\
& WIND-3.5 & \textbf{96.81(2.95)} & 98.67(2.31) & \textbf{5.33(7.57)} & 95.82(2.08) & 97.39(2.09) & 5.88(5.88) \\
& WIND-G   & 94.89(1.35) & 98.67(2.31) & 9.33(4.16) & 96.67(3.34) & \textbf{100.00(0.00)} & 7.15(7.15) \\
& WIND-D   & 96.59(0.44) & \textbf{99.00(1.41)} & 6.00(2.83) & \textbf{96.92(1.41)} & 95.92(0.00) & \textbf{2.04(2.88)} \\
\midrule
\multirow{6}{*}{20}
& BERT     & 96.05(0.79) & 96.02(2.76) & 4.02(3.39) & 93.27(2.68) & 96.75(0.59) & 10.71(8.66) \\
& RoBERTa  & 87.25(3.62) & 90.08(4.66) & 16.41(1.65) & 93.23(1.04) & 93.22(5.36) & 6.75(2.28) \\
& T5       & 85.31(3.44) & 90.76(7.74) & 22.02(7.55) & 85.86(3.87) & 88.02(8.51) & 17.02(2.34) \\
& WIND-3.5 & 97.42(1.95) & 99.33(1.15) & 4.67(5.03) & 97.98(0.02) & \textbf{98.98(1.02)} & 3.02(1.02) \\
& WIND-G   & \textbf{98.05(1.36)} & \textbf{100.00(0.00)} & \textbf{4.00(2.83)} & 97.22(2.21) & 95.92(3.53) & 1.36(1.18) \\
& WIND-D   & 96.74(1.02) & 98.67(2.00) & 5.33(4.16) & \textbf{98.47(0.71)} & 97.96(0.00) & \textbf{1.02(1.44)} \\
\bottomrule
\end{tabular}
\caption{Performance of WIND and baselines on GPT3.5-generated data. Each baseline is fine-tuned with num samples from protected creative writing and negative texts. WIND-3.5, WIND-G, and WIND-D use GPT3.5, Grok, and DeepSeek-V3 \cite{liu2024deepseek} as $G(\cdot)$. Results are reported as means with standard deviations.}
\label{base}
\end{table*}

\section{Experiments}
\label{exp}

\subsection{Dataset and Experimental Setting}
\label{dataset_info}

We use two stylistically distinct human-written datasets as protected creative writing $\bm{T_H}$: Shakespeare (SP) and ROCStories (ROC) \cite{zhu2023storytrans}. Additionally, the IMDB dataset \cite{dai2019style} serves as one of the negative creative works. For example, when the protected set $\bm{T_P}$ consists of LLM-rewritten ROC texts, the negative set $\bm{T_N}$ includes LLM-rewritten SP and IMDB texts. Rewritten texts are generated utilizing two language models: GPT-3.5-turbo-16k (GPT3.5) \cite{brown2020language}, Grok-beta\footnote{https://console.x.ai} (Grok), and OPT-1.3B \cite{zhang2022opt} (OPT). Note that OPT is included solely in the watermark baseline. All tabulated values represent the mean results from three experimental trials. Unless otherwise specified, the default experiment uses 'GPT3.5' to generate imitation texts and 20 samples from the protected and unprotected creative writing, respectively.
Performance is evaluated using True Positive Rate (TPR), False Positive Rate (FPR), and F1 score. We employ SimCSE-RoBERTa \cite{gao2021simcse} as the encoder for our 356.41 M parameter model. WIND is trained with an Apple M1 Pro chip, whose integrated GPU does not expose GPU-level statistics such as memory usage or processing time. We report wall-clock time as a performance proxy; training on the SP dataset with 10 samples takes approximately 26 minutes. WIND is optimized with AdamW \cite{loshchilov2017decoupled}. The learning rate of $E_\alpha(\cdot)$ is tuned between 5e-5 and 1e-7, while $\bm{M}_\gamma$ is fixed at 1e-5. Empirically, we set $mar=0.5$, $\rho=2.0$, $\lambda^+=\lambda^-=\lambda_c=2.0$, and the delimitation threshold to $0.8$. Dataset statistics and implementation details are available in supplements.


\begin{table}[h]
\renewcommand{\arraystretch}{0.45}
\centering
\begin{tabular}{lcccc}
\toprule
\multirow{2}{*}{Method} & \multicolumn{2}{c}{ROC} & \multicolumn{2}{c}{SP} \\
\cmidrule(lr){2-3} \cmidrule(lr){4-5}
& TPR & F1 & TPR & F1 \\
\midrule
KGW       & 97.17 & 95.24 & 93.87 & 92.92 \\
Unigram   & 96.13 & 94.30 & 94.37 & 92.47 \\
EWD       & 88.27 & 91.90 & 93.83 & 94.73 \\
SynthID   & 85.33 & 87.54 & 78.89 & 83.57 \\
Unbiased  & 50.14 & 62.72 & 38.14 & 51.65 \\
MorphMark & 95.31 & 94.72 & 95.73 & 94.28 \\
\midrule
WIND-D    & 98.53 & 97.25 & 98.41 & 96.98 \\
\bottomrule
\end{tabular}
\caption{Comparison of WIND with SOTA Watermarking Methods (FPR below 10\%).}
\label{base_wm10}
\end{table}

\begin{table}[h]
\renewcommand{\arraystretch}{0.45}
\centering
\begin{tabular}{lcccc}
\toprule
\multirow{2}{*}{Method} & \multicolumn{2}{c}{ROC} & \multicolumn{2}{c}{SP} \\
\cmidrule(lr){2-3} \cmidrule(lr){4-5}
& TPR & F1 & TPR & F1 \\
\midrule
KGW       & 88.03 & 93.22 & 89.80 & 94.28 \\
Unigram   & 91.17 & 94.99 & 89.58 & 94.29 \\
EWD       & 88.02 & 93.49 & 91.27 & 95.07 \\
SynthID   & 84.71 & 91.28 & 78.52 & 87.77 \\
Unbiased  & 16.00 & 27.37 & 14.23 & 24.72 \\
MorphMark & 93.48 & 96.53 & 94.87 & 97.19 \\
\midrule
WIND-D    & 97.07 & 98.48 & 97.43 & 98.69 \\
\bottomrule
\end{tabular}
\caption{Comparison of WIND with SOTA Watermarking Methods (FPR below 1\%).}
\label{base_wm1}
\end{table}

\subsection{Main Results}
\label{sub:base}

Our main experiment addresses two key research questions. \textbf{[Q1:] Is a specialized approach like WIND required for creative writing protection?} To explore this, we utilize three powerful fine-tuned classifiers sufficient for a nuanced style protection task, including BERT-base-uncased (BERT) \cite{DBLP:conf/naacl/DevlinCLT19}, RoBERTa \cite{liu2019RoBERTa}, and T5 \cite{raffel2020exploring}. Table \ref{base} presents the results, which two main findings: (1) Overall, WIND surpasses baselines in safeguarding creative writing, while also exhibiting a lower standard deviation. (2) WIND achieves satisfactory F1 scores and minimal FPR with just ten protected style samples, whereas the baselines show unstable performance and often incur high FPR. 

These experiments underscore the necessity of specialized approaches like WIND, verifying unique stylistic signatures. Baseline models incorporate content irrelevant to creativity into their feature space, causing them to prioritize generic content over stylistic essence. This bias leads to protection failures and motivates our proposed instance delimitation mechanism for sample classification. Overall, WIND extracts a creative essence watermark that verifies copyright origin, delivering superior accuracy and reliability.


\textbf{[Q2:] Is WIND superior to state-of-the-art watermarking methods?} 
The baseline SOTA watermarking schemes we compare include: KWG \cite{kirchenbauer2023watermark}, Unigram \cite{zhaoprovable}, EWD \cite{lu-etal-2024-entropy}, SynthID \cite{dathathri2024scalable}, Unbiased \cite{huunbiased}, and MorphMark \cite{wang-etal-2025-morphmark}. Following the original configurations of the baselines, we use OPT-1.3B to generate watermarked creative writing and non-watermarked negative samples. Detailed implementations are provided in supplements. For fairness, WIND's data are also generated by the same model. Tables \ref{base_wm10} and \ref{base_wm1} report results with FPR constrained below 10\% and 1\%, respectively. Experiments reveal that WIND substantially outperforms SOTA methods in validating the creative essence watermark, primarily because our approach condenses creation-specific features into a verifiable and implicit watermark. WIND maintains robust compatibility with detecting suspicious texts without being constrained by infringing models, whether black-box or white-box models.


\subsection{Robustness Study}

We evaluate the robustness of WIND against diverse attack methods. To safeguard creation integrity, attacks must avoid substantial disruptions from creative essence. Attack methods \cite{DBLP:conf/acl/DuganHTZLXIC24}, including case swapping (Upper-Lower), common misspellings (Misspelling), number insertions (Number), adding \textbackslash n\textbackslash n between sentences (Add Paragraph), and utilization of Grok for sentence rewriting with creation (Rewrite), are designed with minimized creation impact. 
Table \ref{robustness} reveals that WIND maintains strong performance even under adversarial attacks, confirming its effectiveness in copyright validation for creative writing.
\begin{table}[h]
\renewcommand{\arraystretch}{0.45}
\centering
\begin{tabular}{lccc}
\toprule
Method & F1 & TPR & FPR \\
\midrule
Upper-Lower   & 97.67$_{\downarrow 0.80}$ & 97.01$_{\downarrow 0.95}$ & 1.64$_{\uparrow 0.62}$ \\
Misspelling   & 97.50$_{\downarrow 0.97}$ & 95.92$_{\downarrow 2.04}$ & 0.84$_{\downarrow 0.18}$ \\
Number        & 97.85$_{\downarrow 0.62}$ & 97.15$_{\downarrow 0.81}$ & 1.42$_{\uparrow 0.40}$ \\
Rewrite       & 95.48$_{\downarrow 2.99}$ & 94.13$_{\downarrow 3.83}$ & 3.04$_{\uparrow 2.02}$ \\
Add Paragraph & 97.13$_{\downarrow 1.34}$ & 96.43$_{\downarrow 1.53}$ & 2.13$_{\uparrow 1.11}$ \\
\midrule
WIND-D        & 98.47 & 97.96 & 1.02 \\
\bottomrule
\end{tabular}
\caption{Robustness study in ROC.}
\label{robustness}
\end{table}

\subsection{Ablation Study} 


We assess each component via an ablation study in Table \ref{abalation}, with five modifications: '$-\mathcal{L}_{con}$' and '$-\mathcal{L}_{style}$' remove contrastive losses in the encoder and condensed-lists, respectively; '$-\mathcal{L}_w$' removes watermark loss; '$-\bm{C}$' skips instance delimitation and LLM condensation; 'Froze $\alpha$' freezes the encoder; and '$-q_p$' sends samples directly to the LLM without delimitation. Removing any component degrades performance. In particular, omitting the delimitation mechanism (row '$-q_p$') reduces F1 by about 9.20\%, confirming its role in providing high-quality input for LLM condensation.


To assess WIND's dependence on prompt quality during LLM condensation, we modify templates by removing illustrative examples and remaining the minimal task description ($q_{low}$). Results show that WIND maintains bit-consistent watermarks within a protected style across prompts. Moreover, BERT and RoBERTa underperform relative to SimCSE-RoBERTa due to reduced model anisotropy.

\begin{table}[h]
\renewcommand{\arraystretch}{0.45}
\centering
\begin{tabular}{lccc}
\toprule
Method & F1 & TPR & FPR \\
\midrule
$-\mathcal{L}_{con}$   & 88.68 & 95.92 & 20.41 \\
$-\mathcal{L}_{style}$ & 94.95 & 95.92 & 6.12  \\
$-\mathcal{L}_{w}$     & 73.21 & 78.05 & 35.17 \\
$-\bm{C}$              & 83.28 & 85.72 & 20.14 \\
Frozen $\alpha$        & 86.88 & 92.53 & 20.48 \\
$q_p$                  & 89.27 & 88.35 & 9.59  \\
$q_{low}$              & 98.17 & 97.83 & 1.48  \\
\midrule
BERT(WIND)             & 92.35 & 93.87 & 9.42  \\
RoBERTa(WIND)          & 94.62 & 95.89 & 6.79  \\
WIND-D                 & 98.47 & 97.96 & 1.02  \\
\bottomrule
\end{tabular}
\caption{Ablation and various encoders results.}
\label{abalation}
\end{table}

\subsection{Further Explorations} 
    
\begin{figure}[h]
    \centering
    \includegraphics[width=2.6in]{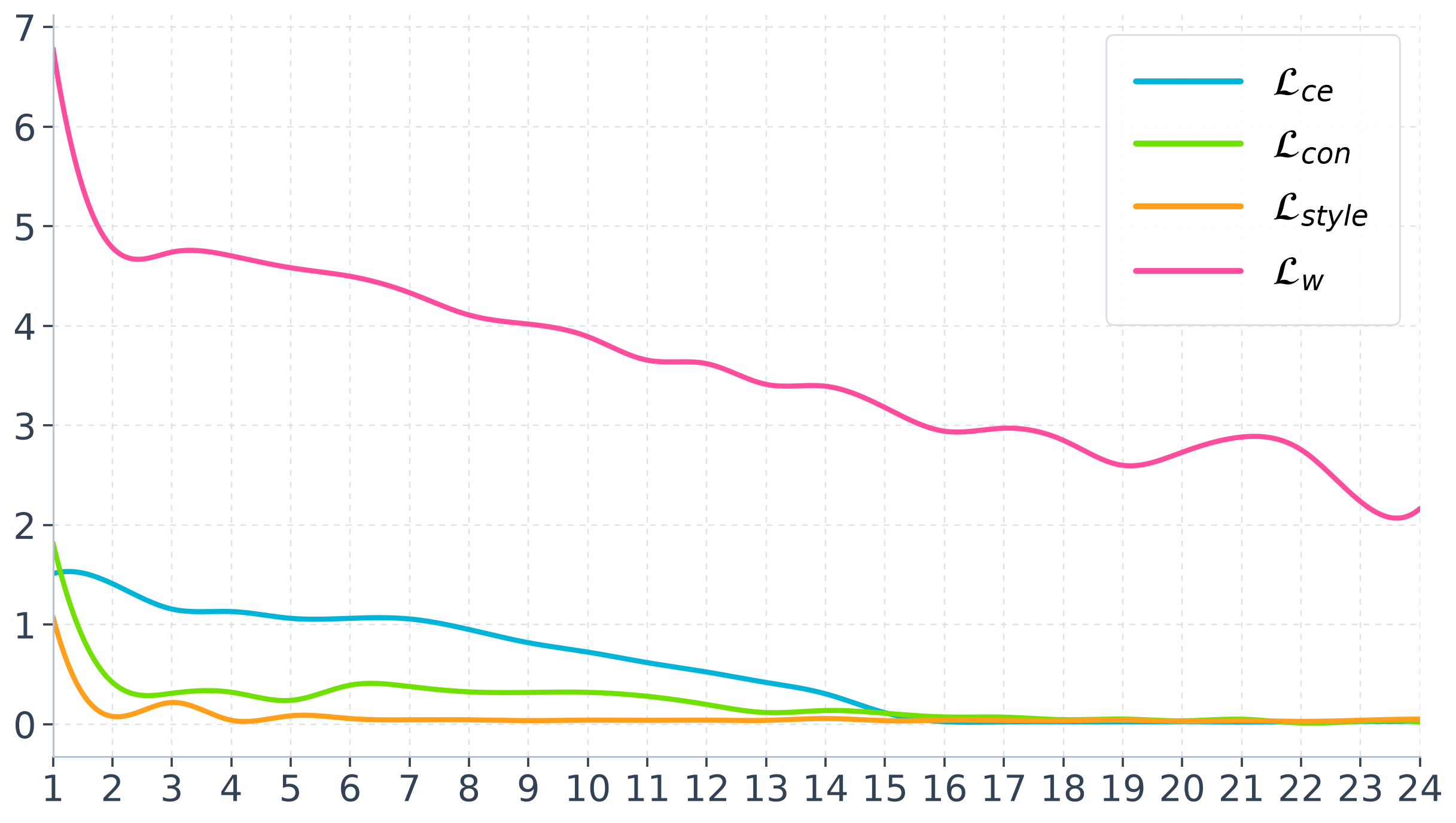}
    \caption{Loss evolution during training.}
    \label{loss_down}
\end{figure}

\textbf{Loss Convergence.} As shown in Figure \ref{loss_down}, all loss components consistently decrease during training:  $\mathcal{L}_{ce}$, $\mathcal{L}_{con}$ and $\mathcal{L}_{style}$ uickly approach zero, while $\mathcal{L}_{w}$ steadily declines across epochs. This convergence empirically validates our core design, showing that the model jointly learns discriminative textual representations, separates condensed stylistic features, and aligns watermarks with the predefined verification target. The stable curves further suggest that these objectives are mutually compatible.

\begin{figure}[h]
    \centering
    \includegraphics[width=2.6in]{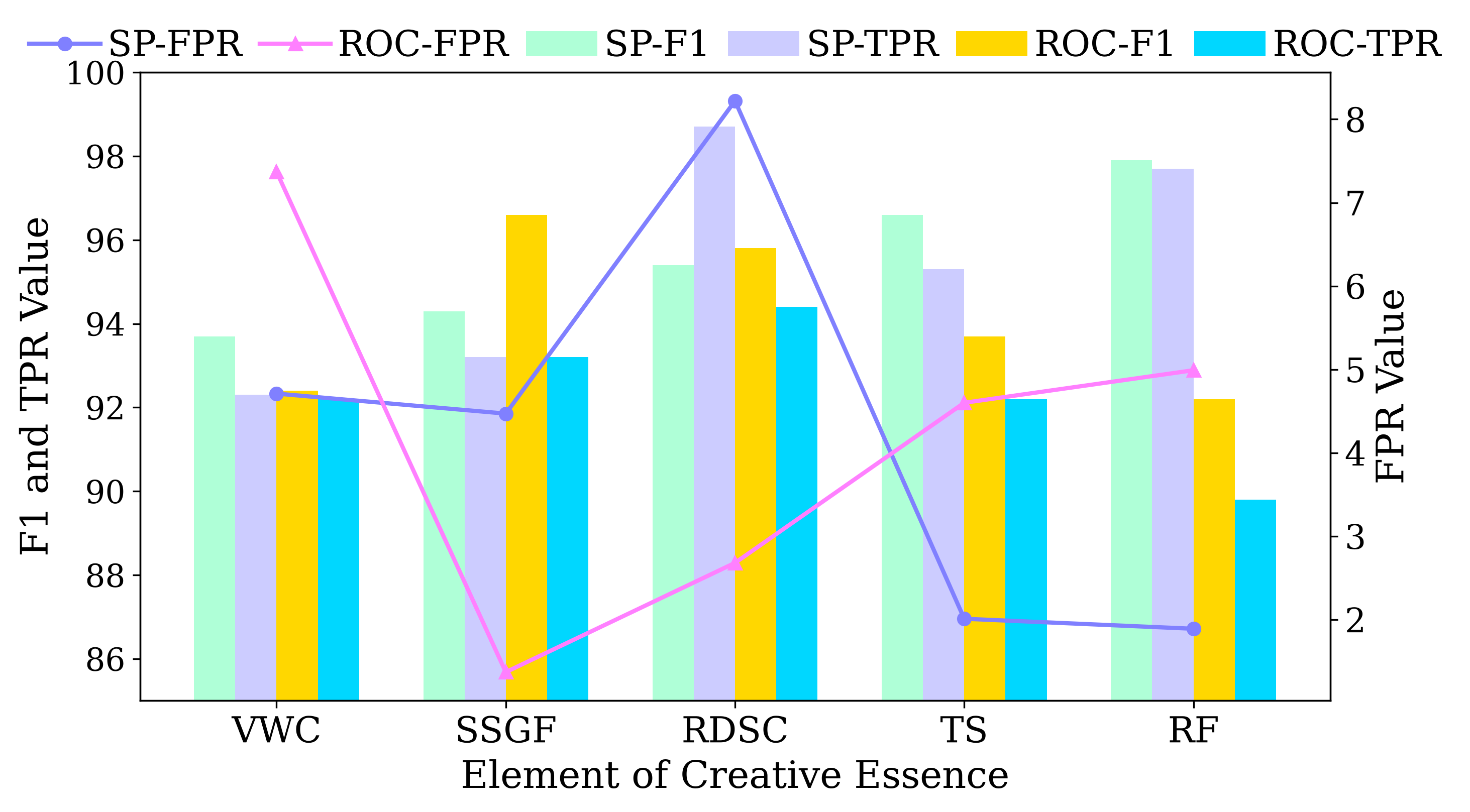}
    \caption{Effectiveness of five elements. (Performance of WIND-D as an illustrative case.)}
    \label{prompt}
\end{figure}

\textbf{Creative element analysis.} Figure \ref{prompt} indicates that omitting specific components leads to performance degradation. This confirms that each element serves as a core attribute of creative essence. For instance, extracting only RF on the ROC dataset significantly decreases performance because RF is less distinctive across the general SP dataset than in poetry. 

To validate the impact of different combinations of creative elements, we test four representative combinations on SP. The results in Table \ref{ele_sen} demonstrate that adding more elements consistently improves performance, confirming their complementary roles. For example, on SP, the 3-element set (VWC+SSGF+TS) outperforms the 2-element one (SSGF+TS). These experiments confirm the efficacy and non-redundancy of the five-element set. Its ability to represent an author's potential multi-genre characteristics allows it to effectively condense the essence of creative writing, achieving comprehensive stylistic coverage where the omission of any component impairs performance.

\begin{table}[h]
\renewcommand{\arraystretch}{0.45}
\centering
\label{ele_sen}
\begin{tabular}{lccc}
\toprule
Method & F1 & TPR & FPR \\
\midrule
SSGF+TS     
& 94.27$_{\downarrow 4.20}$ 
& 94.17$_{\downarrow 3.79}$ 
& 5.62$_{\uparrow 4.60}$ \\

VWC+RF      
& 96.16$_{\downarrow 2.31}$ 
& 93.59$_{\downarrow 4.37}$ 
& 1.06$_{\uparrow 0.04}$ \\

VWC+SSGF+TS 
& 97.21$_{\downarrow 1.26}$ 
& 96.28$_{\downarrow 1.68}$ 
& 1.81$_{\uparrow 0.79}$ \\

RDCS+TS+RF  
& 97.82$_{\downarrow 0.65}$ 
& 97.88$_{\downarrow 0.08}$ 
& 2.24$_{\uparrow 1.22}$ \\

WIND-D      
& 98.47 
& 97.96 
& 1.02 \\
\bottomrule
\end{tabular}
\caption{Results for different combinations of creative elements and prompt sensitive.}
\label{ele_sen}
\end{table}


\textbf{Cross-Model Generalization.} Table \ref{agnostic} summarizes the results of our validation of generalization. The findings demonstrate that WIND maintains consistent performance under a distribution shift caused by training and testing on data from different LLMs.
Supplements provide further exploratory analyses, including baselines when detecting grok-generated data, benchmarking against alternative methods, and presenting a qualitative case study.
\begin{table}[h]
\renewcommand{\arraystretch}{0.45}
    \centering
    \begin{tabular}{c c|c c c}
        \toprule
        $num$ &  & F1 & TPR & FPR \\ \midrule
         & GPT3.5 $\rightarrow$ Grok & 96.75 & 95.01 & 1.39 \\
         10 & Grok $\rightarrow$ GPT3.5 & 95.13 & 94.37 & 4.03 \\
          & WIND-D & 96.92 & 95.92 & 2.04 \\ \midrule
         & GPT3.5 $\rightarrow$ Grok & 97.12 & 96.91 & 2.66 \\
        20   & Grok $\rightarrow$ GPT3.5 & 97.38 & 96.04 & 1.21 \\
           & WIND-D  & 98.47 & 97.96 & 1.02 \\ 
        \bottomrule
    \end{tabular}
        \caption{The notation 'Grok$\rightarrow$GPT3.5' indicates that the model is trained on data generated by Grok but tested on data generated by GPT3.5; the same applies to 'Grok$\rightarrow$GPT3.5'.}
    \label{agnostic}
\end{table}

\section{Conclusion and Limitation}

We introduce WIND, a verifiable and implicit watermarking scheme to enable copyright verification of creative writing against unauthorized AI imitation. First, we decompose the concept of creative essence into five elements and propose an instance delimitation mechanism to guide LLMs in generating condensed lists.
These lists are projected into a disentangled space for watermark mapping. Crucially, WIND preserves creative integrity while operating independently of the infringement model, ensuring broad compatibility with imitation texts. Experimental results demonstrate WIND’s superior practicality and copyright verification robustness.
WIND’s primary limitation is disentanglement performance. Future work will optimize the feature space for creative attributes and integrate style-specific elements to broaden its expressive scope. 


\bibliography{custom}

\begin{thebibliography}{50}
\providecommand{\natexlab}[1]{#1}

\bibitem[{{Baker Hostetler}(2023)}]{getty2023stability}
{Baker Hostetler}. 2023.
\newblock Getty Images v. Stability AI.

\bibitem[{Brown et~al.(2020)Brown, Mann, Ryder, Subbiah, Kaplan, Dhariwal, Neelakantan, Shyam, Sastry, Askell et~al.}]{brown2020language}
Brown, T.; Mann, B.; Ryder, N.; Subbiah, M.; Kaplan, J.~D.; Dhariwal, P.; Neelakantan, A.; Shyam, P.; Sastry, G.; Askell, A.; et~al. 2020.
\newblock Language models are few-shot learners.
\newblock \emph{Advances in neural information processing systems}, 33: 1877--1901.

\bibitem[{Chen et~al.(2022)Chen, Zeng, Ying, Li, Qian, and Zhang}]{9859698}
Chen, K.; Zeng, X.; Ying, Q.; Li, S.; Qian, Z.; and Zhang, X. 2022.
\newblock Invertible Image Dataset Protection.
\newblock In \emph{2022 IEEE International Conference on Multimedia and Expo (ICME)}, 01--06.

\bibitem[{Christ, Gunn, and Zamir(2024)}]{christ2024undetectable}
Christ, M.; Gunn, S.; and Zamir, O. 2024.
\newblock Undetectable watermarks for language models.
\newblock In \emph{The Thirty Seventh Annual Conference on Learning Theory}, 1125--1139. PMLR.

\bibitem[{{Court Listener}(2023)}]{AIauthors}
{Court Listener}. 2023.
\newblock Authors Guild v.OpenAI Inc.(1:23-cv-08292)[DB/OL].(2023-09-19)[2024-04-20].

\bibitem[{Dai et~al.(2019)Dai, Liang, Qiu, and Huang}]{dai2019style}
Dai, N.; Liang, J.; Qiu, X.; and Huang, X.-J. 2019.
\newblock Style Transformer: Unpaired Text Style Transfer without Disentangled Latent Representation.
\newblock In \emph{Proceedings of the 57th Annual Meeting of the Association for Computational Linguistics}, 5997--6007.

\bibitem[{Dathathri et~al.(2024)Dathathri, See, Ghaisas, Huang, McAdam, Welbl, Bachani, Kaskasoli, Stanforth, Matejovicova et~al.}]{dathathri2024scalable}
Dathathri, S.; See, A.; Ghaisas, S.; Huang, P.-S.; McAdam, R.; Welbl, J.; Bachani, V.; Kaskasoli, A.; Stanforth, R.; Matejovicova, T.; et~al. 2024.
\newblock Scalable watermarking for identifying large language model outputs.
\newblock \emph{Nature}, 634(8035): 818--823.

\bibitem[{Devlin et~al.(2019)Devlin, Chang, Lee, and Toutanova}]{DBLP:conf/naacl/DevlinCLT19}
Devlin, J.; Chang, M.; Lee, K.; and Toutanova, K. 2019.
\newblock {BERT:} Pre-training of Deep Bidirectional Transformers for Language Understanding.
\newblock In Burstein, J.; Doran, C.; and Solorio, T., eds., \emph{Proceedings of the 2019 Conference of the North American Chapter of the Association for Computational Linguistics: Human Language Technologies, {NAACL-HLT} 2019, Minneapolis, MN, USA, June 2-7, 2019, Volume 1 (Long and Short Papers)}, 4171--4186. Association for Computational Linguistics.

\bibitem[{Dong et~al.(2024)Dong, Li, Dai, Zheng, Ma, Li, Xia, Xu, Wu, Chang, Sun, Li, and Sui}]{dong-etal-2024-survey}
Dong, Q.; Li, L.; Dai, D.; Zheng, C.; Ma, J.; Li, R.; Xia, H.; Xu, J.; Wu, Z.; Chang, B.; Sun, X.; Li, L.; and Sui, Z. 2024.
\newblock A Survey on In-context Learning.
\newblock In Al-Onaizan, Y.; Bansal, M.; and Chen, Y.-N., eds., \emph{Proceedings of the 2024 Conference on Empirical Methods in Natural Language Processing}, 1107--1128. Miami, Florida, USA: Association for Computational Linguistics.

\bibitem[{Dugan et~al.(2024)Dugan, Hwang, Trhl{\'{\i}}k, Zhu, Ludan, Xu, Ippolito, and Callison{-}Burch}]{DBLP:conf/acl/DuganHTZLXIC24}
Dugan, L.; Hwang, A.; Trhl{\'{\i}}k, F.; Zhu, A.; Ludan, J.~M.; Xu, H.; Ippolito, D.; and Callison{-}Burch, C. 2024.
\newblock {RAID:} {A} Shared Benchmark for Robust Evaluation of Machine-Generated Text Detectors.
\newblock In Ku, L.; Martins, A.; and Srikumar, V., eds., \emph{Proceedings of the 62nd Annual Meeting of the Association for Computational Linguistics (Volume 1: Long Papers), {ACL} 2024, Bangkok, Thailand, August 11-16, 2024}, 12463--12492. Association for Computational Linguistics.

\bibitem[{Feng et~al.(2023)Feng, Ma, Yu, Huang, Wang, Chen, Peng, Feng, Qin et~al.}]{feng2023trends}
Feng, Z.; Ma, W.; Yu, W.; Huang, L.; Wang, H.; Chen, Q.; Peng, W.; Feng, X.; Qin, B.; et~al. 2023.
\newblock Trends in integration of knowledge and large language models: A survey and taxonomy of methods, benchmarks, and applications.
\newblock \emph{arXiv preprint arXiv:2311.05876}.

\bibitem[{Gao, Yao, and Chen(2021)}]{gao2021simcse}
Gao, T.; Yao, X.; and Chen, D. 2021.
\newblock SimCSE: Simple Contrastive Learning of Sentence Embeddings.
\newblock In \emph{Proceedings of the 2021 Conference on Empirical Methods in Natural Language Processing}, 6894--6910.

\bibitem[{Han et~al.(2024)Han, Gao, Liu, Zhang, and Zhang}]{han2024parameter}
Han, Z.; Gao, C.; Liu, J.; Zhang, J.; and Zhang, S.~Q. 2024.
\newblock Parameter-Efficient Fine-Tuning for Large Models: A Comprehensive Survey.
\newblock \emph{Transactions on Machine Learning Research}.

\bibitem[{He et~al.(2022)He, Xu, Lyu, Wu, and Wang}]{he2022protecting}
He, X.; Xu, Q.; Lyu, L.; Wu, F.; and Wang, C. 2022.
\newblock Protecting intellectual property of language generation apis with lexical watermark.
\newblock In \emph{Proceedings of the AAAI Conference on Artificial Intelligence}, volume~36, 10758--10766.

\bibitem[{Hu et~al.(2022)Hu, Shen, Wallis, Allen{-}Zhu, Li, Wang, Wang, and Chen}]{DBLP:conf/iclr/HuSWALWWC22}
Hu, E.~J.; Shen, Y.; Wallis, P.; Allen{-}Zhu, Z.; Li, Y.; Wang, S.; Wang, L.; and Chen, W. 2022.
\newblock LoRA: Low-Rank Adaptation of Large Language Models.
\newblock In \emph{The Tenth International Conference on Learning Representations, {ICLR} 2022, Virtual Event, April 25-29, 2022}. OpenReview.net.

\bibitem[{Hu et~al.(2023)Hu, Chen, Wu, Wu, Zhang, and Huang}]{huunbiased}
Hu, Z.; Chen, L.; Wu, X.; Wu, Y.; Zhang, H.; and Huang, H. 2023.
\newblock Unbiased Watermark for Large Language Models.
\newblock In \emph{The Twelfth International Conference on Learning Representations}.

\bibitem[{Huang et~al.(2024)Huang, Guo, Luo, Qian, Li, and Zhang}]{huangdisentangled}
Huang, J.; Guo, Z.; Luo, G.; Qian, Z.; Li, S.; and Zhang, X. 2024.
\newblock Disentangled Style Domain for Implicit $ z $-Watermark Towards Copyright Protection.
\newblock In \emph{The Thirty-eighth Annual Conference on Neural Information Processing Systems}.

\bibitem[{Kao and Jurafsky(2012)}]{kao2012computational}
Kao, J.; and Jurafsky, D. 2012.
\newblock A computational analysis of style, affect, and imagery in contemporary poetry.
\newblock In \emph{Proceedings of the NAACL-HLT 2012 workshop on computational linguistics for literature}, 8--17.

\bibitem[{Kirchenbauer et~al.(2023)Kirchenbauer, Geiping, Wen, Katz, Miers, and Goldstein}]{kirchenbauer2023watermark}
Kirchenbauer, J.; Geiping, J.; Wen, Y.; Katz, J.; Miers, I.; and Goldstein, T. 2023.
\newblock A watermark for large language models.
\newblock In \emph{International Conference on Machine Learning}, 17061--17084. PMLR.

\bibitem[{Kuditipudi et~al.(2023)Kuditipudi, Thickstun, Hashimoto, and Liang}]{kuditipudirobust}
Kuditipudi, R.; Thickstun, J.; Hashimoto, T.; and Liang, P. 2023.
\newblock Robust Distortion-free Watermarks for Language Models.
\newblock \emph{Transactions on Machine Learning Research}.

\bibitem[{Leidinger, van Rooij, and Shutova(2023)}]{leidinger2023language}
Leidinger, A.; van Rooij, R.; and Shutova, E. 2023.
\newblock The language of prompting: What linguistic properties make a prompt successful?
\newblock In \emph{Findings of the Association for Computational Linguistics: EMNLP 2023}, 9210--9232.

\bibitem[{Liu et~al.(2024{\natexlab{a}})Liu, Feng, Xue, Wang, Wu, Lu, Zhao, Deng, Zhang, Ruan et~al.}]{liu2024deepseek}
Liu, A.; Feng, B.; Xue, B.; Wang, B.; Wu, B.; Lu, C.; Zhao, C.; Deng, C.; Zhang, C.; Ruan, C.; et~al. 2024{\natexlab{a}}.
\newblock Deepseek-v3 technical report.
\newblock \emph{arXiv preprint arXiv:2412.19437}.

\bibitem[{Liu et~al.(2023{\natexlab{a}})Liu, Yuan, Fu, Jiang, Hayashi, and Neubig}]{liu2023pre}
Liu, P.; Yuan, W.; Fu, J.; Jiang, Z.; Hayashi, H.; and Neubig, G. 2023{\natexlab{a}}.
\newblock Pre-train, prompt, and predict: A systematic survey of prompting methods in natural language processing.
\newblock \emph{ACM Computing Surveys}, 55(9): 1--35.

\bibitem[{Liu et~al.(2024{\natexlab{b}})Liu, Qin, Ye, Mou, He, and Wang}]{liu2024adaptive}
Liu, Q.; Qin, J.; Ye, W.; Mou, H.; He, Y.; and Wang, K. 2024{\natexlab{b}}.
\newblock Adaptive Prompt Routing for Arbitrary Text Style Transfer with Pre-trained Language Models.
\newblock In \emph{Proceedings of the AAAI Conference on Artificial Intelligence}, volume~38, 18689--18697.

\bibitem[{Liu et~al.(2022)Liu, Ji, Fu, Tam, Du, Yang, and Tang}]{liu2021p}
Liu, X.; Ji, K.; Fu, Y.; Tam, W.; Du, Z.; Yang, Z.; and Tang, J. 2022.
\newblock {P}-Tuning: Prompt Tuning Can Be Comparable to Fine-tuning Across Scales and Tasks.
\newblock In Muresan, S.; Nakov, P.; and Villavicencio, A., eds., \emph{Proceedings of the 60th Annual Meeting of the Association for Computational Linguistics (Volume 2: Short Papers)}, 61--68. Dublin, Ireland: Association for Computational Linguistics.

\bibitem[{Liu(2019)}]{liu2019RoBERTa}
Liu, Y. 2019.
\newblock Roberta: A robustly optimized bert pretraining approach.
\newblock \emph{arXiv preprint arXiv:1907.11692}, 364.

\bibitem[{Liu et~al.(2023{\natexlab{b}})Liu, Hu, Zhang, and Sun}]{liu2023watermarking}
Liu, Y.; Hu, H.; Zhang, X.; and Sun, L. 2023{\natexlab{b}}.
\newblock Watermarking text data on large language models for dataset copyright protection.
\newblock \emph{arXiv preprint arXiv:2305.13257}.

\bibitem[{Loshchilov and Hutter(2019)}]{loshchilov2017decoupled}
Loshchilov, I.; and Hutter, F. 2019.
\newblock Decoupled Weight Decay Regularization.
\newblock In \emph{International Conference on Learning Representations}.

\bibitem[{Lu(2010)}]{lu2010automatic}
Lu, X. 2010.
\newblock Automatic analysis of syntactic complexity in second language writing.
\newblock \emph{International journal of corpus linguistics}, 15(4): 474--496.

\bibitem[{Lu et~al.(2024)Lu, Liu, Yu, Li, and King}]{lu-etal-2024-entropy}
Lu, Y.; Liu, A.; Yu, D.; Li, J.; and King, I. 2024.
\newblock An Entropy-based Text Watermarking Detection Method.
\newblock In Ku, L.-W.; Martins, A.; and Srikumar, V., eds., \emph{Proceedings of the 62nd Annual Meeting of the Association for Computational Linguistics (Volume 1: Long Papers)}, 11724--11735. Bangkok, Thailand: Association for Computational Linguistics.

\bibitem[{Maini et~al.(2024)Maini, Jia, Papernot, and Dziedzic}]{maini2024llm}
Maini, P.; Jia, H.; Papernot, N.; and Dziedzic, A. 2024.
\newblock LLM Dataset Inference: Did you train on my dataset?
\newblock \emph{Advances in Neural Information Processing Systems}, 37: 124069--124092.

\bibitem[{OpenAI(2023)}]{DBLP:journals/corr/abs-2303-08774}
OpenAI. 2023.
\newblock {GPT-4} Technical Report.
\newblock \emph{CoRR}, abs/2303.08774.

\bibitem[{Pennebaker and King(1999)}]{pennebaker1999linguistic}
Pennebaker, J.~W.; and King, L.~A. 1999.
\newblock Linguistic styles: language use as an individual difference.
\newblock \emph{Journal of personality and social psychology}, 77(6): 1296.

\bibitem[{Raffel et~al.(2020)Raffel, Shazeer, Roberts, Lee, Narang, Matena, Zhou, Li, and Liu}]{raffel2020exploring}
Raffel, C.; Shazeer, N.; Roberts, A.; Lee, K.; Narang, S.; Matena, M.; Zhou, Y.; Li, W.; and Liu, P.~J. 2020.
\newblock Exploring the limits of transfer learning with a unified text-to-text transformer.
\newblock \emph{Journal of machine learning research}, 21(140): 1--67.

\bibitem[{Sahoo et~al.(2024)Sahoo, Singh, Saha, Jain, Mondal, and Chadha}]{sahoo2024systematic}
Sahoo, P.; Singh, A.~K.; Saha, S.; Jain, V.; Mondal, S.; and Chadha, A. 2024.
\newblock A systematic survey of prompt engineering in large language models: Techniques and applications.
\newblock \emph{arXiv preprint arXiv:2402.07927}.

\bibitem[{Salman et~al.(2023)Salman, Khaddaj, Leclerc, Ilyas, and M{\k{a}}dry}]{salman2023raising}
Salman, H.; Khaddaj, A.; Leclerc, G.; Ilyas, A.; and M{\k{a}}dry, A. 2023.
\newblock Raising the cost of malicious AI-powered image editing.
\newblock In \emph{Proceedings of the 40th International Conference on Machine Learning}, 29894--29918.

\bibitem[{Shan et~al.(2023)Shan, Cryan, Wenger, Zheng, Hanocka, and Zhao}]{shan2023glaze}
Shan, S.; Cryan, J.; Wenger, E.; Zheng, H.; Hanocka, R.; and Zhao, B.~Y. 2023.
\newblock Glaze: Protecting artists from style mimicry by $\{$Text-to-Image$\}$ models.
\newblock In \emph{32nd USENIX Security Symposium (USENIX Security 23)}, 2187--2204.

\bibitem[{Shi et~al.(2024)Shi, Ajith, Xia, Huang, Liu, Blevins, Chen, and Zettlemoyer}]{shi2024detecting}
Shi, W.; Ajith, A.; Xia, M.; Huang, Y.; Liu, D.; Blevins, T.; Chen, D.; and Zettlemoyer, L. 2024.
\newblock DETECTING PRETRAINING DATA FROM LARGE LANGUAGE MODELS.
\newblock In \emph{12th International Conference on Learning Representations, ICLR 2024}.

\bibitem[{Steen et~al.(2010)Steen, Dorst, Krennmayr, Kaal, and Herrmann}]{steen2010method}
Steen, G.~J.; Dorst, A.~G.; Krennmayr, T.; Kaal, A.~A.; and Herrmann, J.~B. 2010.
\newblock A method for linguistic metaphor identification.

\bibitem[{Tang et~al.(2023)Tang, Feng, Liu, Yang, and Hu}]{tang2023did}
Tang, R.; Feng, Q.; Liu, N.; Yang, F.; and Hu, X. 2023.
\newblock Did you train on my dataset? towards public dataset protection with cleanlabel backdoor watermarking.
\newblock \emph{ACM SIGKDD Explorations Newsletter}, 25(1): 43--53.

\bibitem[{{The New York Times}(2023{\natexlab{a}})}]{Sarah}
{The New York Times}. 2023{\natexlab{a}}.
\newblock Sarah silverman sues openai and meta over copyright infringement.

\bibitem[{{The New York Times}(2023{\natexlab{b}})}]{Times}
{The New York Times}. 2023{\natexlab{b}}.
\newblock The times sues openai and microsoft over a.i. use of copyrighted work.

\bibitem[{Tweedie and Baayen(1998)}]{tweedie1998variable}
Tweedie, F.~J.; and Baayen, R.~H. 1998.
\newblock How variable may a constant be? Measures of lexical richness in perspective.
\newblock \emph{Computers and the Humanities}, 32: 323--352.

\bibitem[{Vaezi and Rezaei(2019)}]{vaezi2019development}
Vaezi, M.; and Rezaei, S. 2019.
\newblock Development of a rubric for evaluating creative writing: a multi-phase research.
\newblock \emph{New Writing}, 16(3): 303--317.

\bibitem[{Wang et~al.(2024)Wang, Zhu, Liu, Zheng, Chen, and Li}]{wang2024knowledge}
Wang, S.; Zhu, Y.; Liu, H.; Zheng, Z.; Chen, C.; and Li, J. 2024.
\newblock Knowledge editing for large language models: A survey.
\newblock \emph{ACM Computing Surveys}, 57(3): 1--37.

\bibitem[{Wang et~al.(2025)Wang, Gu, Wu, and Yang}]{wang-etal-2025-morphmark}
Wang, Z.; Gu, T.; Wu, B.; and Yang, Y. 2025.
\newblock {M}orph{M}ark: Flexible Adaptive Watermarking for Large Language Models.
\newblock In Che, W.; Nabende, J.; Shutova, E.; and Pilehvar, M.~T., eds., \emph{Proceedings of the 63rd Annual Meeting of the Association for Computational Linguistics (Volume 1: Long Papers)}, 4842--4860. Vienna, Austria: Association for Computational Linguistics.
\newblock ISBN 979-8-89176-251-0.

\bibitem[{Wei et~al.(2022)Wei, Wang, Schuurmans, Bosma, Xia, Chi, Le, Zhou et~al.}]{wei2022chain}
Wei, J.; Wang, X.; Schuurmans, D.; Bosma, M.; Xia, F.; Chi, E.; Le, Q.~V.; Zhou, D.; et~al. 2022.
\newblock Chain-of-thought prompting elicits reasoning in large language models.
\newblock \emph{Advances in neural information processing systems}, 35: 24824--24837.

\bibitem[{Zhang et~al.(2022)Zhang, Roller, Goyal, Artetxe, Chen, Chen, Dewan, Diab, Li, Lin et~al.}]{zhang2022opt}
Zhang, S.; Roller, S.; Goyal, N.; Artetxe, M.; Chen, M.; Chen, S.; Dewan, C.; Diab, M.; Li, X.; Lin, X.~V.; et~al. 2022.
\newblock Opt: Open pre-trained transformer language models.
\newblock \emph{arXiv preprint arXiv:2205.01068}.

\bibitem[{Zhao et~al.(2024)Zhao, Ananth, Li, and Wang}]{zhaoprovable}
Zhao, X.; Ananth, P.~V.; Li, L.; and Wang, Y.-X. 2024.
\newblock Provable Robust Watermarking for AI-Generated Text.
\newblock In \emph{The Twelfth International Conference on Learning Representations}.

\bibitem[{Zhu et~al.(2023)Zhu, Guan, Huang, and Liu}]{zhu2023storytrans}
Zhu, X.; Guan, J.; Huang, M.; and Liu, J. 2023.
\newblock StoryTrans: Non-Parallel Story Author-Style Transfer with Discourse Representations and Content Enhancing.
\newblock In \emph{Proceedings of the 61st Annual Meeting of the Association for Computational Linguistics (Volume 1: Long Papers)}, 14803--14819.

\end{thebibliography}


\end{document}